\documentclass{aa}
\usepackage[T1]{fontenc}
\usepackage{times}
\usepackage{graphicx}
\usepackage{aabib99}

\newcommand{\parth}[3]{\left(\frac{#1}{#2\,\mathrm{#3}}\right)}
\newcommand{\be}{\begin{equation}}
\newcommand{\ee}{\end{equation}}
\begin{document}

\thesaurus{13.07.1; 02.13.1; 02.18.5; 02.19.1}
\title{Large scale magnetic fields and their dissipation in GRB fireballs}
\mail{henk@mpa-garching.mpg.de}
\author{H. C. Spruit \and 
F. Daigne \and
G. Drenkhahn}
\institute{Max-Planck-Institut f\"ur Astrophysik, Postfach 1317, 85741
  Garching bei M\"unchen, Germany}
\date{Received ... / Accepted ...}
\maketitle

\begin{abstract}
  We consider possible geometries of magnetic fields in GRB outflows,
  and their evolution with distance from the source. For magnetically
  driven outflows, with an assumed ratio of magnetic to kinetic energy
  density of order unity, the field strengths are sufficient for
  efficient production of $\gamma$-rays by synchrotron emission in the
  standard internal shock scenario, without the need for local
  generation of small scale fields. In these conditions, the MHD
  approximation is valid to large distances ($\ga 10^{19}$\,cm). In
  outflows driven by nonaxisymmetric magnetic fields, changes of
  direction of the field cause dissipation of magnetic energy by
  reconnection. This dissipation takes place outside the photosphere
  of the outflow, and can convert a significant fraction of the
  magnetic energy flux into radiation.

\end{abstract}

\section{Introduction}

Several models for cosmic gamma--ray bursts (hereafter GRBs) make use
of rapidly rotating compact stellar-mass sources. Though many details
in each case are uncertain, the two more developed and popular
scenarios involve the coalescence of two compact objects (neutron star
+ neutron star or neutron star + black hole) and the collapse of a
massive star to a black hole (collapsar)
\cite{narayan:92,woosley:93,paczynski:98}. Both lead to the same
system: a stellar mass black hole surrounded by a thick torus made of
stellar debris or of infalling stellar material partially supported by
centrifugal forces. An other interesting proposition
\cite{usov:92,kluzniak:98,spruit:99} associates GRBs with highly
magnetized millisecond pulsars.

The release of energy by such a source can origin from various
reservoirs. In the case of a thick disk + black hole system, the burst
can be powered by the accretion of disk material by the black hole or
by extracting directly the rotational energy of the black hole via the
Blandford-Znajek mechanism. In the case of a highly magnetised
millisecond pulsar the energy release comes from the rotational energy
of the neutron star.

Luminosities as high as those of GRBs cannot be radiated in the close
vicinity of the source. The energy released must first drive a wind
which rapidly becomes relativistic. Its kinetic energy is then
converted into gamma-rays (producing the prompt emission of the GRB)
at large radii via the formation of shocks, probably within the wind
itself (internal shock model) \cite{rees:94,daigne:98}. The wind is
finally decelerated by the external medium, which leads to a new shock
responsible for the afterglow emission observed in X--rays, optical
and radio bands (external shock model) \cite{wijers:97}.

The Lorentz factor $\Gamma$ of the relativistic wind must reach high
values ($\Gamma \sim 10^{2}$--$10^{3}$) both to produce gamma--rays
and to avoid photon--photon annihilation along the line of sight,
whose signature is not observed in the spectra of GRBs as observed by
\textit{BATSE} (see e.g. Baring 1995\nocite{baring:95}). This limits
the allowed amount of baryonic pollution in the flow to a very low
level and makes the production of the relativistic wind a challenging
problem.  Only a few ideas have been proposed and none appears to be
fully conclusive at present.

However it is suspected that large magnetic fields play an important
role. This is obvious in models using a highly magnetised millisecond
pulsar, but in the case of a thick disk + black hole system a magnetic
wind is also probably more efficient than the initial proposition
where the wind is powered by the annihilation of
neutrino--antineutrinos pairs along the rotational axis
\cite{ruffert:97}.  The magnetic field in the disk could be amplified
by differential rotation to very large values ($B \sim 10^{15}$\,G)
and a magnetically driven wind could then be emitted from the disk
\cite{thompson:94,meszaros:97}. Under severe constraints on the field
geometry and the dissipation close to the disk, large values of the
Lorentz factor can then be reached \cite{daigne:00a,daigne:00b}.

The emission of photons at large radii via the formation of shocks is
perhaps better understood than the central engine. It is believed that
a non--thermal population of accelerated electrons is produced behind
the shock waves and synchrotron and/or inverse Compton photons are
emitted. A magnetic field is then required. In the case of the
afterglow emission, the external shock is propagating in the
environment of the source and the magnetic field has to be locally
generated by microscopic processes
\cite{meszaros:93a,wijers:97,thompson:00}. In the case of the prompt
gamma--ray emission which is probably due to internal shocks within
the wind, such a locally generated magnetic field is also usually
invoked \cite{rees:94,papathanassiou:96,sari:97,daigne:98}. It has
also been pointed out that a large scale field originating from the
source could play the same role \cite{meszaros:93b,meszaros:97}.

The argumentation in this paper is organized as follows. In
Sect.~\ref{sec:mhd} we show that for typical baryon loading, the
particle density in the outflow is so large that the MHD approximation
is appropriate out to distances of the order $10^{20}$\,cm. 
This will turn out to be larger than the other relevant distances. 

Due to the baryon loading, the GRB case is therefore different from the
essentially baryon-free pulsar wind case (e.g. \cite{usov:94}), where the 
MHD approximation breaks down much earlier, and plasma theories of 
large amplitude electromagnetic waves (LAEMW) are applied. The GRB
case is actually simpler than the pulsar case in this respect.

The evolution
of the magnetic field can therefore be dealt with in MHD
approximation. In Sect.~\ref{sec:geo} we discuss how the strength and
configuration of the field in the outflow depends on assumptions about
the central engine. This is done first without allowing for decay of
the field by internal MHD processes. In Sect.~\ref{sec:intshock} we
show that the field strengths so obtained are sufficient to produce
synchrotron and/or inverse Compton emission in the standard internal
shock model, without the need for local generation of microscopic
magnetic fields.  In Sect.~\ref{sec:decay} we then argue that internal
MHD processes are, in fact, likely to cause magnetic field energy
density to be released during the outflow.  This may be a significant
contributor to the observed emission.  The efficiency of conversion of
the primary energy of the central engine to observable gamma rays can
also be much higher than in internal shock models. The arguments are
summarized again in Sect.~\ref{sec:disc}.


\section{The validity of the MHD approximation}
\label{sec:mhd}

In this section, as well as throughout this paper, a prime denotes
quantities measured in the rest frame of the outflowing matter,
unprimed quantities are measured in the `laboratory' frame (understood
here as a frame at rest relative to the central object of the burst).

In order to maintain the necessary current there must be enough plasma
available.  As an example, consider the case of a magnetized outflow
with magnetic field of alternating direction, as happens in a
pulsar-like model (Sect.~\ref{sec:qs}).  We approximate this as a
plane sinusoidal wave with a wave number $k$ and a angular frequency
$\Omega$:
\begin{eqnarray}
  \label{eq:waves}
  \vec E &=& E \sin(kz-\Omega t) \vec e_x\\
  \vec B &=& B \sin(kz-\Omega t) \vec e_y.
\end{eqnarray}
In a frame comoving with the fluid the field strength is
time-independent, hence the phase speed of the wave in the lab frame
is $\Omega/k=\beta c$.  From the induction equation $\nabla\times\vec
E=-\partial_t\vec B/c$ the relation between the electrical and
magnetic field amplitudes is $E= \beta B$.  Amp\`ere's equation gives
the current density, $\vec j = -\Omega/(4\pi) B(\beta^{-1}-\beta)
\cos(kz-\Omega t)\vec e_x$, or, with $\Gamma=(1-\beta^2)^{-1/2}$:
\begin{equation}
  \label{eq:j}
  j = \frac{ckB}{4\pi\Gamma^2}.
\end{equation}
(The same result is obtained if the current density is calculated in
MHD approximation in the comoving frame and then Lorentz transformed
back into the lab frame.) The current that can be maintained by the
outflowing medium is limited by the density of charged particles $n$.
The maximum current density (in the comoving frame) is
$j'_\mathrm{m}=n'ec$, when the particles are given their maximum speed
$c$. In the lab frame this current density is \cite{melatos:96}
\begin{equation}
  \label{eq:jm}
  j_\mathrm{m}=nec/\Gamma.
\end{equation}
If $j$ gets larger than $j_\mathrm{m}$ the simple ansatz of an
electromagnetic wave traveling with the same speed as the plasma
cannot hold any longer.  Thus flux freezing is impossible and MHD is
not sufficient any more to describe the problem.

Plasma physical instabilities can set in at current densities much
lower than (\ref{eq:jm}). They will produce an anomalous resistivity
in the plasma so that an electric field is present also in the rest
frame of the plasma. The electric field due to this resistivity is
small, however, compared to the magnetic field strength as long as the
charged particle density is larger than the minimum density. The MHD
approximation in this sense is then a good one on large scales, in
spite of the presence of small scale plasma processes.

If the outflow has a finite duration $\tau$, and a constant speed
$\beta c$ in this time interval, it moves as a shell of width $\beta c
\tau\approx c\tau$. When the shell (assumed spherical) has expanded to
a radius $r$, the particle density measured in the lab frame is $n(r)
= M/(2\pi r^2 c\tau m_\mathrm{p})$ where $M$ is the total mass ejected
and $m_\mathrm{p}$ the proton mass.
With (\ref{eq:j}) and (\ref{eq:jm}), the minimum particle density
$n_\mathrm{c}$ is
\begin{equation}
  \label{eq:nc}
  n_\mathrm{c}(r)=\frac{k B}{4\pi e\Gamma}.
\end{equation}
If internal MHD processes can be neglected, the total electro-magnetic
energy $E_\mathrm{em}$ of the shell is conserved (see
Sect.~\ref{sec:geo}).  The field strength in the lab frame is then
\begin{equation}
   B=\sqrt{\frac{2 E_\mathrm{em}}{c\tau r^2}}.
\end{equation}

With (\ref{eq:nc}), and the known density $n(r)$, we thus find that
the MHD approximation breaks down when the shell has reached the
critical radius
\begin{equation}
  \label{eq:rc}
  r_\mathrm{c}
  = \frac{e E_\mathrm{k}}
  {\Omega m_\mathrm{p}c}
  \sqrt{\frac{2}{E_\mathrm{em}\tau c}}
  = \frac{e}{\Omega m_\mathrm{p}c}
  \sqrt{\frac{2}{\xi(\xi+1)}\frac{E}{c\tau}}
\end{equation}
where $E_\mathrm{k} = \Gamma Mc^2$ is the kinetic energy, $\xi =
E_\mathrm{em}/E_\mathrm{k}$ (in the lab frame) and $E =
E_\mathrm{em}+E_\mathrm{k}$ is the total energy.
Assuming $\xi$ to be of the order unity we have
\begin{equation}
  r_\mathrm{c}\approx
  \frac{e}{\Omega m_\mathrm{p}c}\sqrt{\frac{E}{c\tau}}
  = 3.2\cdot 10^{20}\,\mathrm{cm}\cdot 
  E_{52}^{1/2}
  \Omega_4^{-1}
  \parth{\tau}{3}{s}^{-1}.
\end{equation}
The processes discussed in this paper happen well within time scales
smaller than $r_\mathrm{cr}/c$, and the MHD is justified.

In baryon-free cases, the expected values of $\xi$ may be much higher,
as in the typical pulsar-wind scenarios (e.g. Blackman \& Yi
1998\nocite{blackman:98}), and the critical radius for MHD condition
correspondingly smaller. Such low baryon loading seems rather
unlikely, however, for the currently proposed scenarios for GRB
engines. Outflows from accretion disks, merging stars, and supernova
envelopes are all intrinsically highly baryon loaded environments, and
have some difficulties reaching baryon loadings as small as $\eta\sim
10^{-3}-10^{-4}$. Even in the rapidly magnetized msec neutron star
scenario \cite{spruit:99}, emergence of magnetic fields from the star
at the required short time scales is likely to imply that some
baryonic mass is lifted together with the magnetic fields.


\section{Field geometries}
\label{sec:geo}

\begin{figure}
  \hfill\includegraphics[width=0.17\hsize]{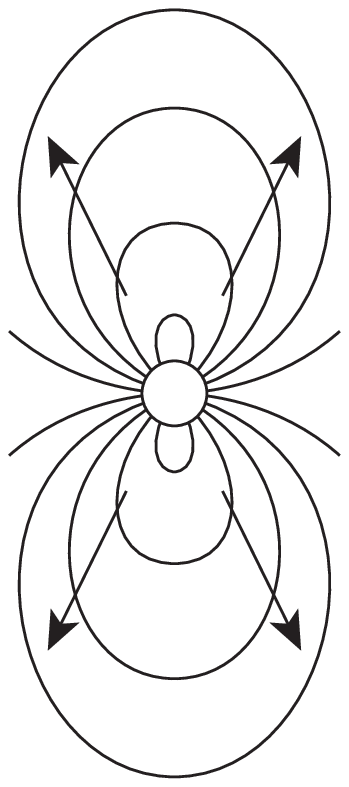}\hfill
  \includegraphics[width=0.40\hsize]{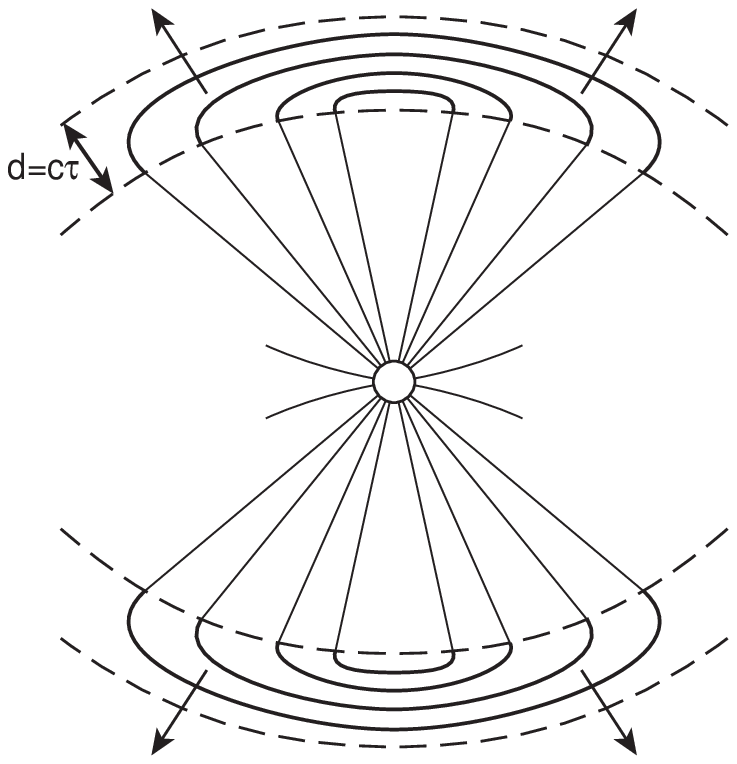}\hfill\mbox{}\\[0.5cm]
  \mbox{}\hfill\includegraphics[width=0.27\hsize]{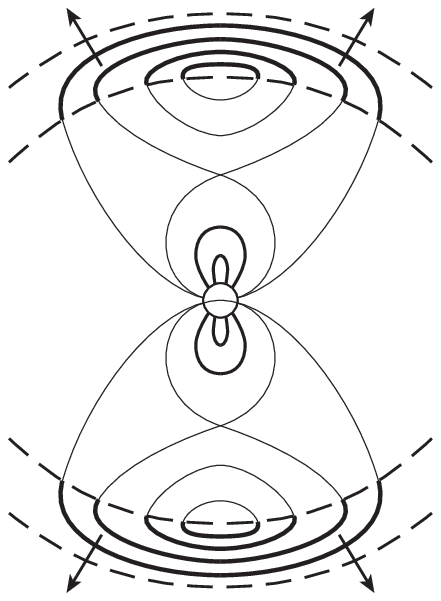}\hfill
  \includegraphics[width=0.50\hsize]{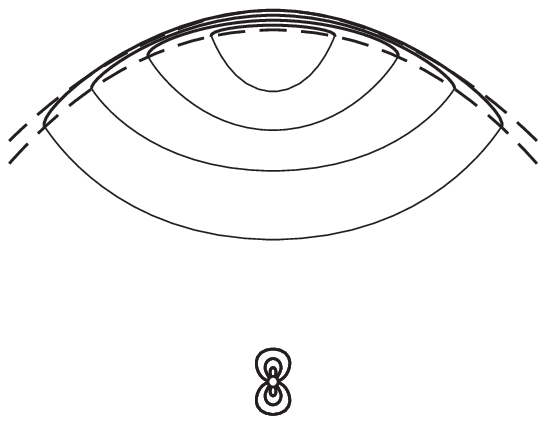}\hfill\mbox{}
  \caption{Evolution of an initial dipole field by a radial outflow 
    (schematic).  Top left: initial field configuration. Top right:
    view on a larger scale, when the outflow, of finite duration
    $\tau$, now moves in a shell of width $c\tau$. It has stretched
    the field interior to the shell into a radial field. Lower left:
    reconnection in the low-density interior region restores the
    dipole field near the source. Lower right: as the shell moves out
    its thickness becomes small compared the distance traveled, and
    the further evolution depends on reconnection processes inside the
    shell.}
  \label{fig1} 
\end{figure}

A burst of duration $\tau$ produces a shell of width $d=c\tau$,
traveling outward at a speed $\beta c$. The observed radiation from
this shell is produced when it has expanded to a radius $r\gg d$. This
thin shell carries with it magnetic field lines from the source, which
we call `trapped field lines'. For the rest of this section, we assume
a relativistic outflow, $\beta\approx 1$.

\subsection{Trapped fields}

If reconnection processes inside the shell can be neglected, the field
lines are frozen in the expanding shell. If $\vec B = (B_r, B_\theta,
B_\phi)$ (in spherical coordinates), and the width of the shell is
constant, the components then vary with distance as $B_r \sim r^{-2}$,
$B_\theta \sim B_\phi\sim r^{-1}$. This is because the radial
component is divided over the surface of the spherical shell, while
the components parallel to the shell surface decrease with the
circumference. More formally, the induction equation $\partial_t \vec
B = \nabla\times(\vec u\times\vec B)$ has the components
\begin{eqnarray}
  \label{eq:ind}
  \partial_t B_\theta&=& (r\sin\theta)^{-1} \partial_\phi 
  (u_\theta B_\phi - u_\phi B_\theta)\nonumber\\
  &&- r^{-1} \partial_r(r (u_r B_\theta - u_\theta B_r))\\
  \partial_t B_\phi&=& r^{-1} (\partial_r(r 
    (u_\phi B_r - u_r B_\phi))\nonumber\\
    &&- \partial_\theta (u_\theta B_\phi 
    - u_\phi B_\theta)).
\end{eqnarray}
Assuming constant radial outflow $(u_r=\mathrm{const.}, u_\theta =
u_\phi=0)$ and spherical symmetry $(\partial_\theta = \partial_\phi =
0)$ the time evolution in a comoving fluid element is
\begin{equation}
  \label{eq:ind2}
  \frac{\mathrm{d} B_{\theta,\phi}}{\mathrm{d}t}
  =\partial_t B_{\theta,\phi} + u_r \partial_r B_{\theta,\phi}
  =-r^{-1} \frac{\mathrm{d}r}{\mathrm{d}t} B_{\theta,\phi}.
\end{equation}
Hence the tangential field $\vec B_\mathrm{t} = (B_\theta, B_\phi)
\sim r^{-1}$. Since the expansion factor is very large between the
source and the radius from which the emitted radiation reaches us, the
radial component (varying as $\sim r^{-2}$)can be neglected and the
field is almost exactly parallel to the surface of the shell
(Fig.~\ref{fig1}). If the width of the shell is constant, the magnetic
energy in the shell $e_\mathrm{m} = \int B_\mathrm{t}^2 \mathrm{d}S$
is constant. The magnetic field thus carries a \emph{constant
  fraction} of the kinetic energy of the outflow.  Depending on how
large this fraction is, the trapped field can be sufficient to produce
the synchrotron emission proposed in internal shock models, without
the need for `in situ' field generation mechanisms \cite{medvedev:99}.

How many field lines are trapped in the outflow, and hence which
fraction of the outflow energy is magnetic, depends on conditions near
the source.  We consider here the representative possibilities.

\subsection{A `passive' magnetic field}

First consider a fireball expanding from the surface of a central
object (not specified further) of radius $R$ and a dipolar magnetic
field of moment $\mu$. We assume that the magnetic field plays no role
in the driving of the outflow, but that it can confine plasma up to
its own energy density $B^2/8\pi$.

Assume first that the expansion is kinematic, i.e. that the magnetic
field is weak and its backreaction on the flow can be neglected. The
number of field lines trapped in the outflow is equal to the flux of
field lines outside crossing the dipole's equator outside the radius
$R$:
\begin{equation}
  \Phi=\frac{\pi\mu}{R}.
\end{equation}
This flux is a Lorentz invariant, so that the field strength of the
shell (as measured in the frame of the central engine) is of the order
$B=\Phi/(2\pi rc\tau)$, or
\begin{equation}
  \label{eq:Bpass}
  B\approx B_0\frac{R^2}{rc\tau}
\end{equation}
where $B_0=\mu/R^3$.  Whether this is a large field strength, compared
with the kinetic energy density $\Gamma\rho c^2$, depends on the
assumed dipole moment $\mu$. An upper limit to the dipole moment
follows from the requirement that the energy of the burst should be
able to open the field lines of the central object. Let the energy of
the burst be $\Gamma Mc^2$, where $\Gamma$ the asymptotic Lorentz
factor of the outflow and $M$ the baryon load. If this energy was
initially confined in a region of size $R$ (the size of the central
engine), the magnetic field strength $B_0$ in the confining dipole
field of the source must satisfy
\begin{equation}
  \frac{B_0^2}{8\pi} < \frac{\Gamma Mc^2}{\frac{4}{3}\pi R^3}
\end{equation}
in order for the field to be opened up by the fireball. With
(\ref{eq:Bpass}), the magnetic energy density at distance $r$ then
satisfies
\begin{equation}
  \label{eq:bpas}
  \frac{B^2}{8\pi} < e_\mathrm{k} \frac{R}{3c\tau},
\end{equation}
where $e_\mathrm{k}$ is the kinetic energy density in the shell:
\begin{equation}
  e_\mathrm{k}=\frac{\Gamma Mc^2}{4\pi r^2c\tau}.
\end{equation}
For most central engine models considered, the duration of the burst
is long compared to the light travel time across the source,
$R/c\tau\ll 1$.  A `passive' model, in which the magnetic field does
not play a role in driving the outflow, therefore can only yield field
strengths in the shell which are small compared with kinetic energy
density. Even at such a low field strength, however, the magnetic
field can become important for synchrotron emission in internal
shocks, as discussed below in Sect.~\ref{sec:intshock}.

\subsection{Active magnetic fields}

In magnetic models for GRB engines, the magnetic field serves to
extract rotation energy from a rapidly rotating relativistic object.
The details of such magnetic extraction (especially three-dimensional
ones) are still somewhat uncertain, but basic energetic considerations
are simple. Rotation of the mass-loaded field lines induces an
azimuthal field component $B_\phi$. Let the distance from the rotating
object where this component is equal to the radial field be $r_0$.
Except for cases with large baryon loading that are probably not
relevant for GRB engines, $r_0$ is of the order of the Alfv\'en
radius, which can lie anywhere between the surface of the object $R$
and the light cylinder $r_\mathrm{L}=c/\Omega$. The energy output $L$
transmitted by the azimuthal magnetic stress ($B_\phi B_r/4\pi$) is
then of the order $L\approx\Omega r_0^3B_r^2(r_0)$. For a field
dominated by its dipole component $\mu$, this yields a luminosity
\begin{equation}
  \label{eq:lum}
  L=\Omega\mu^2/r_0^3.
\end{equation}
For $r_0=r_\mathrm{L}$, this yields (by order of magnitude) the pulsar
spin-down formula for an inclined dipole rotating in vacuum, emitting
an electromagnetic wave from its light cylinder.

\begin{figure}
  \parbox[c]{0.5\hsize}{
    \includegraphics[width=\hsize]{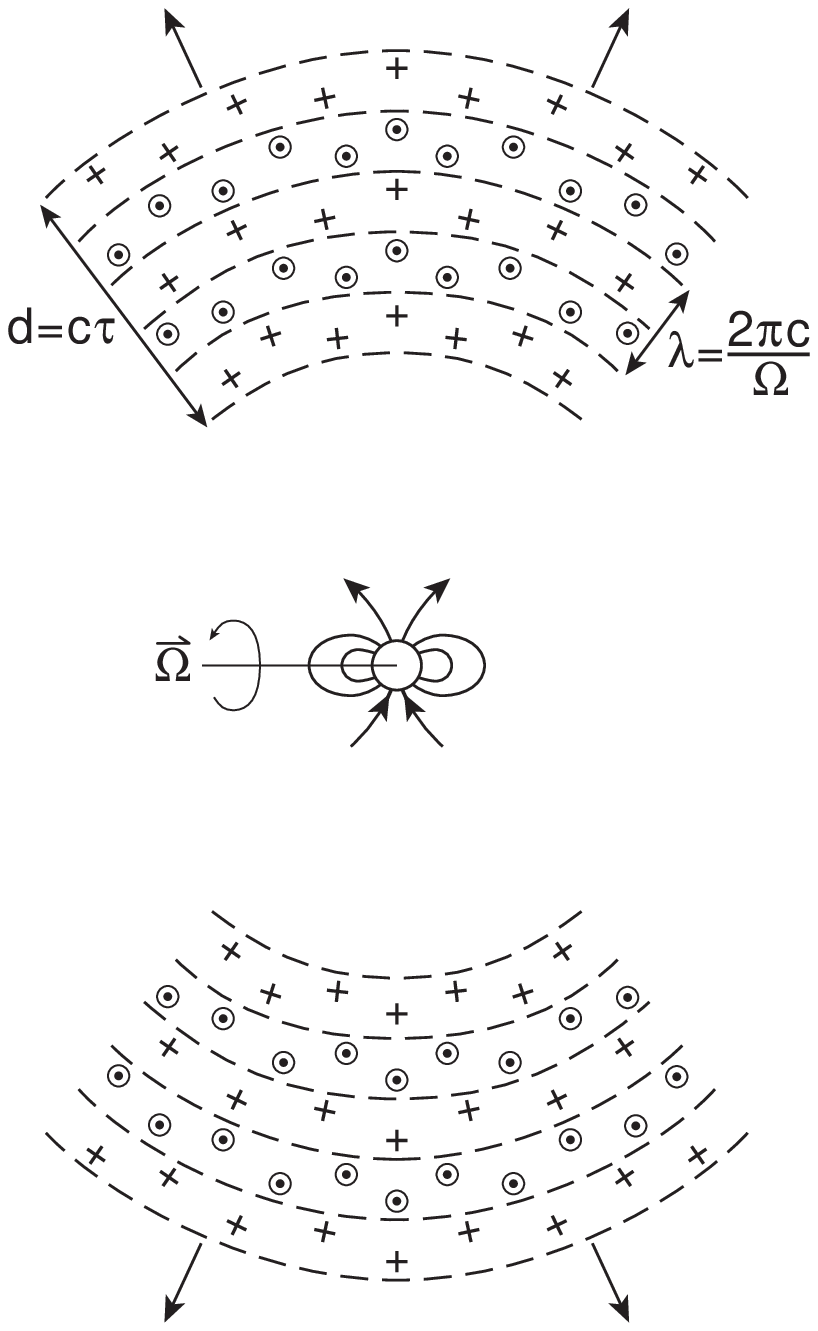}
  }
  \hfill
  \parbox[c]{0.4\hsize}{
    \begin{center}
      \includegraphics[width=\hsize]{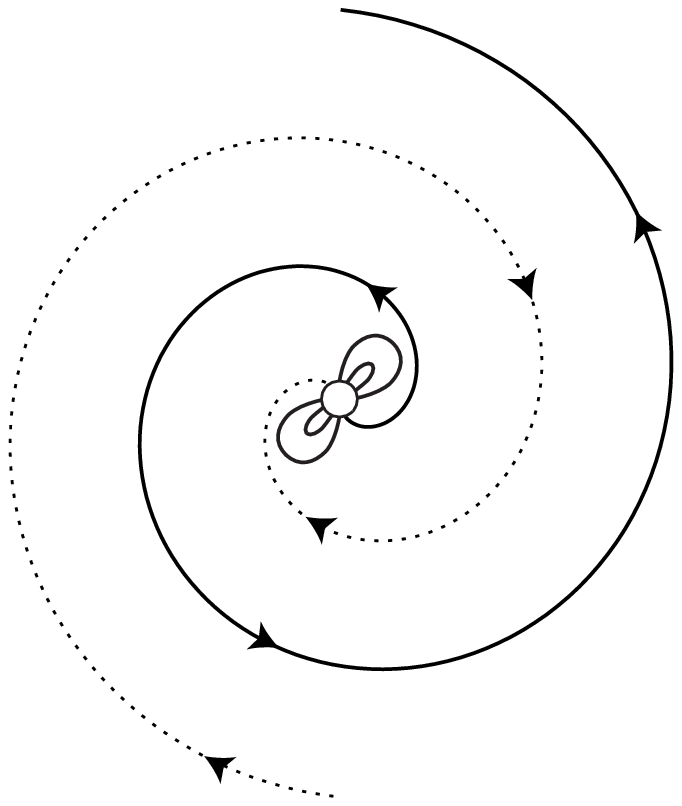}\\[1ex]
      \includegraphics[width=\hsize]{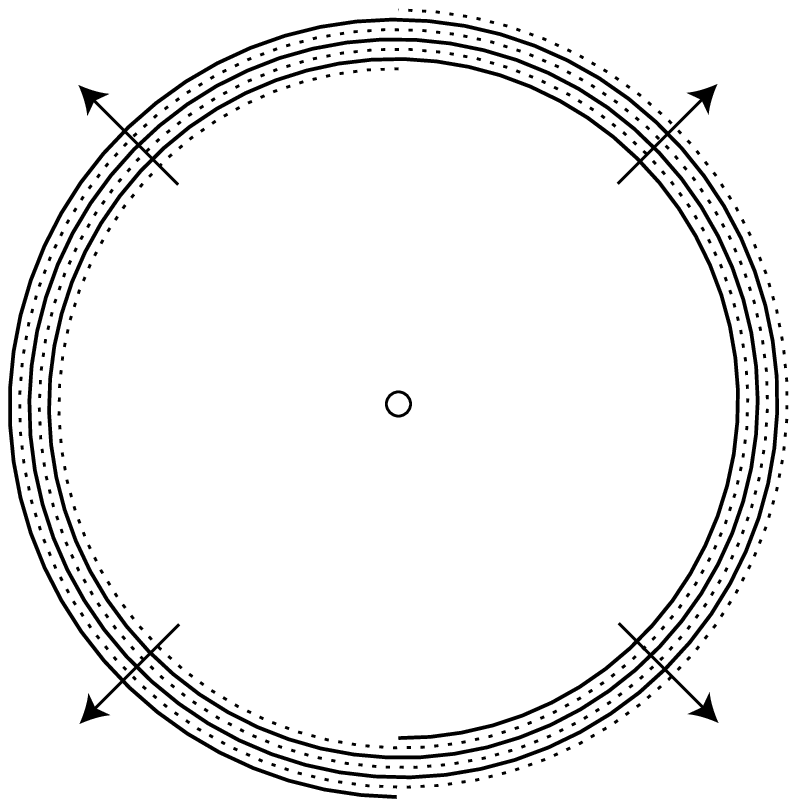}  
    \end{center}
  }
  \caption{Field configuration in quasi-spherical magnetic outflow 
    driven by a perpendicular rotator (`pulsar-like' case)
    (schematic). Left: view in the equatorial plane, with dots and
    plusses indicating field lines into and out of the plane of the
    drawing. Right: top view from the rotational pole. Bottom right:
    same view on larger scale, at a later time $t\gg\tau$.}
  \label{fig2}
\end{figure}

The energy estimate (\ref{eq:lum}) does not tell what the field
configuration in the outflow looks like. The possibilities for
dissipation of magnetic energy inside the outflow depend strongly on
this configuration, which in turn depends on the nature of the nature
of the magnetic field of the central engine. An important distinction
is whether the rotating magnetic field is \emph{axisymmetric} or
\emph{nonaxisymmetric}. In the nonaxisymmetric case, the outflow
contains magnetic fields varying on the rather small scale $\pi
c/\Omega$, the wavelength of the outgoing wave. In such a field,
internal dissipation turns out to be much more likely to be important
than in an axisymmetric field, where the length scale of the field is
of the order of the distance $r$. In the following, this is
illustrated with a few specific cases.

\subsubsection{ratio of magnetic to kinetic energy}

Estimate (\ref{eq:lum}) only gives the total luminosity. Which
fraction of it is in the form of kinetic energy and which in magnetic
energy depends on more physics. At one extreme is Michel's
\cite*{michel:69} model, in which the outflow consists of cold
(pressureless) matter accelerated exclusively by the magnetic field.
In this case, the ratio $\alpha = E_\mathrm{m}/E_\mathrm{k}$ of
magnetic to kinetic luminosity is of the order $\xi\sim \Gamma^2\gg
1$. At large Lorentz factors, the energy is almost entirely in
electromagnetic form, in this model. This is probably a property of
the special geometry of the model, in which the flow is limited to the
equatorial plane. If outflow in other directions is considered, much
smaller values of $\xi$ result \cite{begelman:94}.
  
At the other extreme, consider a case in which most magnetic energy is
released inside the light cylinder, in the form of a dense pair plasma
(by some form of magnetic reconnection, for example). The result would
then be just like the hydrodynamic expansion of a simple
\cite{paczynski:86} pair plasma fireball, and the resulting outflow
would have $\xi\ll 1$. If some but not all energy is dissipated close
to the source, intermediate values of $\xi$ would result.  Since these
important questions have not been resolved yet, we keep $\xi$ as a
free parameter in the following. Where necessary we assume that it
typically has values or order unity.

\subsection{Nonaxisymmetric quasi-spherical outflow}
\label{sec:qs}

Assume we have a perpendicular rotator, i.e. the rotating object has
its dipole axis orthogonal to the rotation axis. At the source surface
$r_0$, the rotating field is then strongest at the equator, and one
expects the energy flux to be largest near the equatorial plane. With
each rotation, a `stripe' consisting of a band of eastward and one of
westward azimuthal field moves outward (Fig.~\ref{fig2}, see also
Coroniti 1990; Usov 1994\nocite{coroniti:90,usov:94}). Assuming the
outflow to be relativistic, the width $\lambda$ of such a stripe in
the rest frame of the central object is $\lambda = 2\pi c/\Omega$. The
(absolute value of the) azimuthal flux $\Phi$ in each half-wavelength
is of the order of the poloidal flux outside the source surface $r_0$,
$\Phi\approx 2\pi\mu/r_0$.  For spherical expansion of this amount of
azimuthal flux, the field strength at distance $r$ is then
\begin{equation}
  B_\phi\approx\frac{\Phi}{r\lambda}
  =\frac{\mu}{r_0 r_\mathrm{L} r}
\end{equation}
while the the total (magnetic plus kinetic energy) density
$e_\mathrm{k}$) is
\begin{equation}
  e_\mathrm{k}+e_\mathrm{m}=\frac{L}{4\pi r^2c}.
\end{equation}
Hence with (\ref{eq:lum}) the ratio of magnetic energy density to
total energy density (in the lab frame) is of the order
\begin{equation}
  \label{eq:beqr}
  e_\mathrm{m}/(e_\mathrm{k}+e_\mathrm{m})
  \approx\frac{r_0}{r_\mathrm{L}}.  
\end{equation}

\subsection{Jet}
\label{sec:jet}

\begin{figure}
  \includegraphics[height=6.5cm]{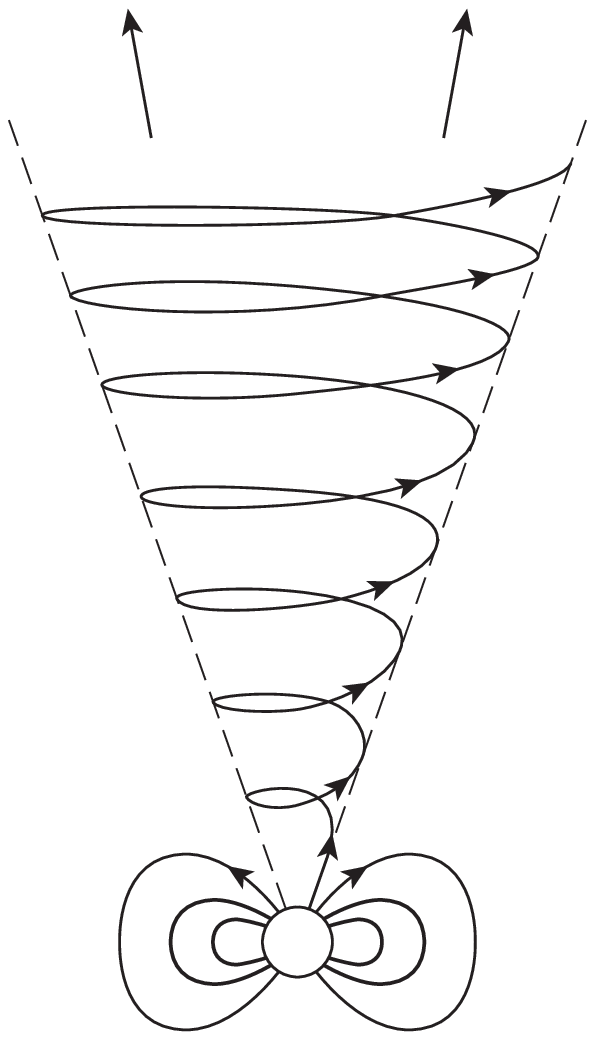}\hfill
  \includegraphics[height=6.5cm]{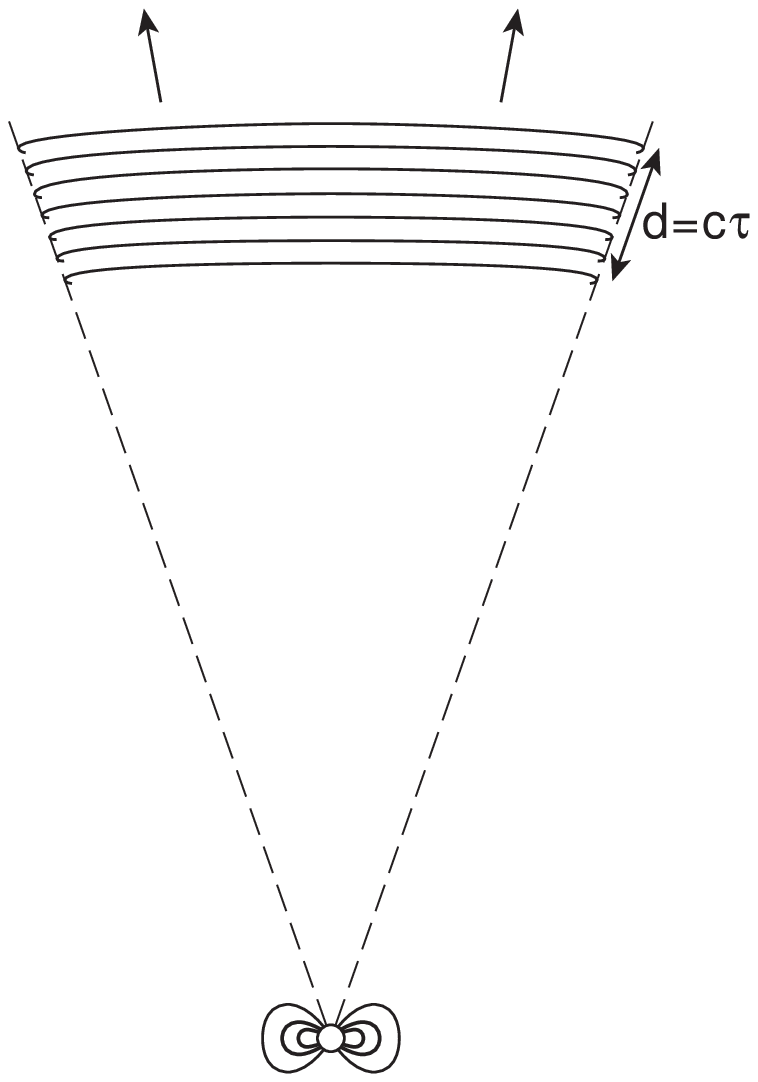}
  \caption{Jet-like outflow of finite duration, magnetically 
    driven by an axially symmetric rotator. Left: Configuration near
    the source, showing how the field in the outflow is wound into a
    toroidal (azimuthal) field. Right: large scale view after a time
    long compared to the duration $\tau$. The field in the outflow is
    now a `pancake' of toroidal flux. Sketch ignores nonaxisymmetric
    processes like kink instability and subsequent reconnection, which
    can release energy from this configuration}
  \label{fig3}
\end{figure}

Consider a collimated outflow along the axis of rotation
(Fig.~\ref{fig3}). This might be achieved by magnetic models in which
the azimuthal field collimates the flow towards the axis by hoop
stress (e.g. Sakurai 1985\nocite{sakurai:85}). In such a model, one
assumes an axisymmetric (about the rotation axis) poloidal field,
which is wound up into an azimuthal field wrapped around the axis. Let
the opening angle of the outflow be $\theta$ (assumed constant), and
$\varpi$ the cylindrical radius. At the source surface $\varpi_0 =
\theta z_0$ the poloidal and azimuthal field components are equal, and
in the absence of magnetic reconnection processes
\begin{eqnarray}
  B_\phi&=&
  B_\mathrm{p0}(\varpi_0/\varpi)=
  B_\mathrm{p0}(z_0/z),\\
  B_\mathrm{p}&=&
  B_\mathrm{p0}(z_0/z)^2.
\end{eqnarray}
If the collimation angle is small, the field as seen in a frame
comoving with the jet is a slowly varying, nearly azimuthal field.
Such a field is known to be highly unstable to kink instabilities
(e.g. Bateman 1980\nocite{bateman:80}). They operate on a time scale
$\tau_\mathrm{k}$ of the order of the Alfv\'en crossing time, i.e.
$\tau_\mathrm{k} = \varpi/v_\mathrm{A}$. Though the details of this
process have not been worked out for jets (see however Lucek \& Bell
1996, 1997\nocite{lucek:96,lucek:97}), it is likely that the release
of magnetic energy operates in two steps. In the first step, kink
instability transforms the axisymmetric configuration into a
nonaxisymmetric, helical, configuration. For an application to jets,
see K\"onigl \& Choudhuri \cite*{koenigl:85}.  This step is fast,
operating on the Alv\'en crossing time. At this stage, the field has
already lost much of its collimating ability, since the average
azimuthal field strength has decreased in favor of a less ordered
field component. During the instability, the so-called magnetic
helicity is conserved, however, so that only a fraction of the
magnetic energy is released. The further release of magnetic energy
depends on reconnection processes. As discussed in
Sect.~\ref{sec:decay}, this is also likely to proceed on a time scale
proportional to the Alfv\'en crossing time, though somewhat slower
than the kink process itself.
    
As in the quasi-spherical outflow case, we ignore this internal
dissipation of the magnetic energy for the moment, and return to it in
Sect.~\ref{sec:decay}.  For a jet expanding with fixed opening angle
$\theta$ (see Fig.~\ref{fig3}), the field strength then varies as
$r^{-1}$ and the magnetic energy is constant with distance. Since only
a fraction of the magnetic energy is released in the kinking process,
the ratio of magnetic to kinetic energy density is still of the order
found in axisymmetric magnetic jet calculations (e.g.  Camenzind
1987\nocite{camenzind:87}), i.e. of order unity:
\begin{equation}
  \label{eq:bjet}
  e_\mathrm{m}/e_\mathrm{k}\sim O(1)
\end{equation}

\section{The emission expected from internal shocks}
\label{sec:intshock}

\subsection{Typical parameters of an internal shock}

The internal shock model assumes that the initial distribution of the
Lorentz factor in the shell is highly variable. Rapid layers catch up
with slower ones leading to internal shocks propagating within the
relativistic shell. The hot material behind the shock waves is
radiating efficiently and produces the observed prompt gamma--ray
emission of GRBs
\cite{rees:94,kobayashi:97,daigne:98,panaitescu:99,daigne:00c}. We
only summarize here the basic assumptions of the model. For details
see Daigne \& Mochkovitch \cite*{daigne:98}.

Consider two layers of equal mass (for simplicity) and of Lorentz
factor $\Gamma_1$ and $\Gamma_2$ ($\Gamma_1>\Gamma_2$) which are
emitted on a timescale $t_\mathrm{var}$. They will collide at a radius
\begin{eqnarray}
  r_\mathrm{IS} 
  & \simeq & 
  \frac{2\Gamma_{1}^{2}\Gamma_{2}^{2}}{\Gamma_{1}^{2}-\Gamma_{2}^{2}} 
  c t_\mathrm{var} \sim \Gamma^{2} c t_\mathrm{var}\\
  & \sim & 3\cdot 10^{15}\,\mathrm{cm}\cdot
  \left(\frac{\Gamma}{300}\right)^{2}
  \parth{t_\mathrm{var}}{1}{s},
\end{eqnarray}
where $\Gamma$ is as usual the mean Lorentz factor of the flow.  The
average energy which is dissipated per proton in this shock is given
by
\begin{equation}
  \epsilon = \left(\Gamma_\mathrm{int}-1\right) m_\mathrm{p} c^{2}
\end{equation}
with 
\begin{equation}
  \Gamma_\mathrm{int} =
  \frac{1}{2}\left[\sqrt{\frac{\Gamma_1}{\Gamma_2}}
    +\sqrt{\frac{\Gamma_2}{\Gamma_1}}\right].
\end{equation}
With $\Gamma_1 / \Gamma_2 = 4$, which corresponds to a midly
relativistic shock ($v_\mathrm{rel} = 0.88 c$), this gives $\epsilon
\sim 240$\,MeV.

It is generally assumed
\cite{rees:94,papathanassiou:96,sari:97,daigne:98} that behind the
shock wave a fraction of the electrons come into (at least partial)
equipartition with the protons and become highly relativistic. If a
fraction $\alpha_\mathrm{e}$ of the dissipated energy goes into a
fraction $\zeta$ of the electrons, their characteristic Lorentz factor
(in the comoving frame) will be
\begin{equation}
  \Gamma'_\mathrm{e} \simeq \frac{\alpha_\mathrm{e}}{\zeta}
  \frac{\epsilon}{m_\mathrm{e} c^{2}},
\end{equation}
which, for complete equipartition and $\Gamma_1/\Gamma_2 \sim 4$,
yields $\Gamma'_\mathrm{e} \sim 100$ -- $200$ if $\zeta \sim 1$ and
$\Gamma'_\mathrm{e} \ga 10^{4}$ -- $2\cdot 10^{4}$ if only a small
fraction of the electrons are accelerated ($\zeta \la 0.01$. For some
theoretical arguments in favour of such an assumption, see e.g.  Bykov
\& M\'esz\'aros \cite*{bykov:96}). Such highly relativistic electrons
can emit gamma--rays by the synchrotron and/or the inverse Compton
process. A magnetic field is however required.

\subsection{Locally generated versus large--scale magnetic field}

The magnetic field involved in the synchrotron radiation is usually
assumed to be generated locally by microscopic processes. Such a
process has been proposed by Medvedev \& Loeb \cite*{medvedev:99}.
(This is also assumed for the external shock propagating in the
interstellar medium and responsible for the afterglow; see however
Thompson \& Madau \cite*{thompson:00}). If this magnetic field is also
into equipartition with the protons and electrons, it will have
typical values of
\begin{equation}
  B'_\mathrm{eq} \simeq \sqrt{8\pi\ \alpha_\mathrm{eq} n' \epsilon},
\end{equation}
where $\alpha_\mathrm{eq} \le 1/3$ and $n'$ is the comoving proton
number density, which can be estimated by
\begin{eqnarray}
  n' & \simeq & 
  \frac{E_\mathrm{k}}{4\pi r^{2} \Gamma^{2} c^{3} m_\mathrm{p} \tau}\\
  &  \simeq & 2.0\cdot 10^7\,\mathrm{cm^{-3}}\cdot
  r_{15}^{-2}
  E_\mathrm{k,52}
  \left(\frac{\Gamma}{300}\right)^{-2}
  \tau_1^{-1}.
  \label{eq:RhoTyp}
\end{eqnarray}
For typical radii $r\sim 10^{15}$--$10^{16}$\,cm and
$\alpha_\mathrm{eq} \simeq \alpha_\mathrm{e} \simeq 1/3$, this leads
to $B'_\mathrm{eq} \sim 100$--$1000$\,G depending on the contrast
$\Gamma_1/\Gamma_2$.

If the GRB is powered magnetically, however, the outflow is naturally
magetized. (One good reason for assuming magnetic powering is that
alternatives like $\nu\bar\nu$ annihilation are energetically
inefficient).  As shown in Sect.~\ref{sec:geo}, the magnetic energy
content of the outflow is constant as long as internal dissipation of
magnetic energy can be neglected.  For the three cases considered, the
ratio $\alpha_\mathrm{LS}$ of the (large scale-) magnetic to kinetic
energy density is
\begin{eqnarray}
  \alpha_\mathrm{LS}&\equiv& B^2/(8\pi e_\mathrm{K})\sim R/(c\tau),\\
  \alpha_\mathrm{LS}&\sim& r_0/r_\mathrm{L},\\
  \alpha_\mathrm{LS}&\sim& O(1),
\end{eqnarray}
for a passively expanded source field, a magnetically driven
quasi-spherical outflow and a magnetically driven collimated jet
respectively [cf. Eqs. (\ref{eq:bpas}, \ref{eq:beqr}, \ref{eq:bjet})].
For source sizes $R = 10^6$--$10^7$\,cm and durations $\tau =
0.3$--$30$\,s the passively expanding field case has
$\alpha_\mathrm{LS} \sim 10^{-6}$--$10^{-3}$. The magnetically driven
quasi-spherical outflow would yield higher values,
$\alpha_\mathrm{LS}\sim 0.1-1$, assuming $r_0/r_\mathrm{L}\sim 0.1-1$.
A magnetically driven jet would have a $\alpha_\mathrm{LS}\sim 1$.
[These values hold as long as internal dissipation of magnetic energy
can be ignored (Sect.~\ref{sec:decay})].

The corresponding comoving magnetic field in all cases is
\begin{equation}
  B'_\mathrm{LS} \sim 
  \frac{1}{\Gamma} \sqrt{8\pi \alpha_\mathrm{LS} e_K} 
  \sim \sqrt{8\pi \alpha_\mathrm{LS} n' m_\mathrm{p} c^{2}}
\end{equation}
which leads to (using the estimation of $n'$ given by
Eq.~(\ref{eq:RhoTyp})):
\begin{equation}
  B'_{\mathrm LS} \simeq
  500\,\mathrm{G}\cdot 
  \alpha_\mathrm{LS}^{1/2}
  r_{15}^{-1}
  E_\mathrm{k,52}^{1/2}
  \left(\frac{\Gamma}{300}\right)^{-1}
  \tau_1^{-1/2}.
\end{equation}
We see that
\begin{equation}
  \frac{B'_\mathrm{eq}}{B'_\mathrm{LS}} \simeq
  \left(\frac{\alpha_\mathrm{eq}}{\alpha_\mathrm{LS}}\right)^{1/2} 
  \frac{\epsilon}{m_\mathrm{p} c^{2}}
\end{equation}
and that, depending on the values of $\alpha_\mathrm{eq}$ and
$\alpha_\mathrm{LS}$, the large scale magnetic field naturally has
strengths comparable with the equipartition fields usually invoked.

\subsection{Synchrotron and Inverse--Compton emission}

Synchrotron emission by accelerated electrons in a magnetic field
occurs at a typical energy (observer frame)
\begin{equation}
  E_\mathrm{syn} = 1.7\cdot 10^{-8}\,\mathrm{eV}\cdot
  \Gamma B'
  \Gamma^{\prime 2}_\mathrm{e}.
\end{equation}

Consider first the case where a large fraction of the electrons takes
part in the acceleration process, $\zeta \sim 1$. If the equipartition
is complete -- equal amounts of energy in protons and electrons (and
magnetic fields if locally generated) -- we then have
$\Gamma'_\mathrm{e} \sim 50$--$500$ so that for $B' \sim
100$--$1000$\,G the synchrotron photon energy
\begin{equation}
  E_\mathrm{syn} = 
  50\,\mathrm{eV}\cdot
  \left(\frac{\Gamma}{300}\right) 
  B'_3
  \Gamma^{\prime 2}_\mathrm{e,2}
\end{equation}
is in the UV band. Gamma--rays can then be produced by inverse Compton
scattering on the synchrotron photons. We are in the Thomson limit
where
\begin{equation}
  w = \frac{\Gamma'_\mathrm{e} E'_\mathrm{syn}}{m_\mathrm{e} c^{2}} 
  \simeq 3\cdot 10^{-5}\cdot
  B'_3
  \Gamma^{\prime 3}_\mathrm{e,2}
  \ll 1
\end{equation}
so that
\begin{eqnarray}
  E_\mathrm{IC} &\simeq&
  {\Gamma'_\mathrm{e}}^{2} E_\mathrm{syn}\\
  &\simeq&
  500\,\mathrm{keV}\cdot
  \left(\frac{\Gamma}{300}\right)
  B'_3
  \Gamma^{\prime 4}_\mathrm{e,2}.
\end{eqnarray}
The fraction of the electron energy which is radiated in the
gamma--ray range has been computed by Daigne \& Mochkovitch
\cite*{daigne:98} and is given by
\begin{equation}
  \alpha_\mathrm{IC} \simeq 
  \frac{\tau_* {\Gamma'_{e}}^{2}}{1+\tau_* {\Gamma'_{e}}^{2}},
\end{equation}
where the optical $\tau_*$ depth of the layer of relativistic
electrons is the solution of the equation
\begin{eqnarray}
  \tau_{*} {\Gamma'_{e}}^{2} \left( 1+ \tau_{*} {\Gamma'_{e}}^{2}\right)
  & \simeq & 
  0.3\cdot 
  \frac{\alpha_\mathrm{e}}{\alpha_\mathrm{eq}}\\
  & \simeq & 
  0.06\cdot
  \parth{\epsilon}{200}{MeV}
  \frac{\alpha_\mathrm{e}}{\alpha_\mathrm{LS}},
\end{eqnarray}
for an equipartition field and a large--scale magnetic field
respectively.  Low values of the magnetic field increase the
efficiency $\alpha_\mathrm{IC}$ but $B'$ cannot be too small because
(i) $E_\mathrm{IC}$ must stay in the gamma--ray range and (ii) the
typical time--scale of the radiative process $t'_\mathrm{rad}$ must be
shorter than the expansion time $t'_\mathrm{ex}$, otherwise the hot
layer is cooling adiabatically.  These two time--scales are
\begin{equation}
  t'_\mathrm{ex} \simeq
  \frac{r}{\Gamma c} \simeq 
  100\,\mathrm{s}\cdot
  r_{15}
  \left(\frac{\Gamma}{300}\right)^{-1}
\label{eq:tex}
\end{equation}
and
\begin{eqnarray}
  t'_\mathrm{rad} 
  &\simeq& \frac{t'_\mathrm{syn}}{1+\tau_{*} {\Gamma'_{e}}^{2}}\\
  &\simeq& \frac{6\,\mathrm{s}}{1+\tau_{*} {\Gamma'_{e}}^{2}}
  B^{\prime -2}_3
  \Gamma^{\prime -1}_{e,2}.
\end{eqnarray}
The radiative efficiency (energy loss by radiation versus adiabatic
cooling) is then of the order
\begin{equation}
  f_\mathrm{rad} \simeq 
  \frac{t'_\mathrm{ex}}{t'_\mathrm{ex}+t'_\mathrm{rad}}.
\end{equation}

We see that (i) for the equipartition magnetic field these two
conditions limit $\alpha_\mathrm{eq} / \alpha_\mathrm{e}$ to the range
$0.01$--$1$ yielding low total efficiencies, $\alpha_\mathrm{IC}
\times f_\mathrm{rad} \sim 0.4$--$0.1$; (ii) for a passively expanded
source field ($\alpha_\mathrm{LS} \sim 10^{-6}$--$10^{-3}$) it is
impossible to produce gamma--rays for the lowest values of
$\alpha_\mathrm{LS}$ (both because $E_\mathrm{IC}$ is too low and
$f_\mathrm{rad}$ is very small). For $\alpha_\mathrm{LS} \sim 10^{-3}$
the magnetic field is still weak but for short bursts ($n'$ is higher)
or high contrasts of Lorentz factors ($\epsilon$ is higher) it is
possible to produce a gamma--ray burst with a total efficiency
$\alpha_\mathrm{IC} \times f_\mathrm{rad} \sim 0.6$ which is then
quite high; (iii) for a magnetically driven quasi-spherical outflow
($\alpha_\mathrm{LS}\sim 0.1$--$1$) the radiative efficiency is high
($f_\mathrm{rad} \sim 1$), and there is no difficulty to produce
gamma--rays via the inverse Compton emission.
However the efficiency of the IC process becomes very low if
$\alpha_\mathrm{LS}$ is to close to unity ($\alpha_\mathrm{IC} \sim
0.02$ for $\alpha_\mathrm{LS} = 1$) and is still low
($\alpha_\mathrm{IC} \sim 0.2$) for $\alpha_\mathrm{LS} = 0.1$. Values
of $\epsilon$ larger than $200$\,MeV does not improve this efficiency
much.  (iv) For a magnetically driven jet
the situation is very close to the
previous case with $\alpha_\mathrm{LS}\sim 0.1$ and the IC efficiency
is then very low.

We consider now the other extreme case where only a small fraction of
the electrons are accelerated : $\zeta \la 0.01$. The Lorentz factor
of the electrons then reaches very high values of $5000$--$50000$ and
gamma--rays can be produced directly by synchrotron emission:
\begin{equation}
  E_\mathrm{syn} \simeq 
  500\,\mathrm{keV}\cdot
  \left(\frac{\Gamma}{300}\right)
  B'_3
  \Gamma^{\prime 2}_\mathrm{e,4}.
\end{equation}
We are is the Klein--Nishina limit where
\begin{equation}
  w = \frac{\Gamma'_\mathrm{e} E'_\mathrm{syn}}{m_\mathrm{e} c^{2}} 
  \simeq 30\cdot
  B'_3
  \Gamma^{\prime 3}_\mathrm{e,4}
  \gg 1.
\end{equation}
The fraction of the energy which is directly radiated by synchrotron
photons is now given by
\begin{equation}
  \alpha_\mathrm{syn} \simeq 
  \frac{\tau_{*} {\Gamma'_{e}}^{2}/w}{1+\tau_{*} {\Gamma'_{e}}^{2}/w}
\end{equation}
and the optical depth is solution of a more complexe equation (the
opacity has decreased compared to the Thomson regime by a factor
depending on $w$) :
\begin{eqnarray}
  \frac{\tau_{*} {\Gamma'_{e}}^{2}}{w} 
  \left( 1+ \frac{\tau_{*} {\Gamma'_{e}}^{2}}{w}\right)
  & \simeq & 
  0.3\cdot \frac{\alpha_\mathrm{e}}{\alpha_\mathrm{eq}} 
  \frac{3}{8 w^{2}}\left(1 + \ln{\left(2 w \right)}\right)\\
  & \simeq & 
  0.06 \parth{\epsilon}{200}{MeV}\nonumber\\
  & & 
  \cdot \frac{\alpha_\mathrm{e}}{\alpha_\mathrm{LS}}
  \frac{3}{8 w^{2}}\left(1 + \ln{\left(2 w \right)}\right),
\end{eqnarray}
for the equipartition and large--scale magnetic field respectively. As
before we also need to check that the radiative time--scale
\begin{equation}
  t'_\mathrm{rad} 
  \simeq \frac{0.06\,\mathrm{s}}{1+\tau_{*} {\Gamma'_e}^2/w}
  B^{\prime -2}_3
  \Gamma^{\prime -1}_{e,4}.
\end{equation}
is shorter than $t'_\mathrm{ex}$. This is now always the case
($f_\mathrm{rad} \sim 1$) except for the lower values of
$\alpha_\mathrm{LS}$ ($10^{-6}$) or for the last internal shocks
occuring far from the source.

We then can conclude that (i) for an equipartition field
$\alpha_\mathrm{eq}$ has to be of the same order as
$\alpha_\mathrm{e}$ in order for the synchrotron process is to produce
gamma--rays at high efficiency ($\alpha_\mathrm{syn} > 0.9$). (ii) For
a passively expanded source field the case $\alpha_\mathrm{LS} \sim
10^{-6}$ is again excluded because of an extremely low efficiency and
a typical energy which is more in the X--rays range. The case
$\alpha_\mathrm{LS} \sim 10^{-3}$ suffers the same limitations as
before, the efficiency is only $\alpha_\mathrm{syn} \sim 0.1$. (iii)
For a magnetically driven quasi-spherical outflow and a magnetically
driven jet the situation is similar to the equipartition case : the
efficiency is very high ($\alpha_\mathrm{syn} > 0.9$ for a
quasi-spherical outflow and $\alpha_\mathrm{syn} > 0.6$ for a jet).

In conclusion for this section, a passive expanded source field is
probably too weak in most cases to produce a gamma--ray burst without
a locally generated magnetic field, but in the two other cases
described in this paper, a magnetically driven quasi-spherical outflow
and a magnetically driven collimated jet, there is no difficulty to
produce gamma--rays without any need of a supplementary field.  The
efficiencies of the radiative process in these magnetically driven
cases are comparable to those already calculated for an equipartition
field. In particular, as was already pointed out in Daigne \&
Mochkovitch \cite*{daigne:98}, this efficiency is expected to be
higher if the gamma--rays are directly produced by synchrotron
emission (which is possible if only a fraction of the electrons are
accelerated behind the shock waves).


\section{Magnetic energy release in the outflow}
\label{sec:decay}

Internal release of magnetic energy can be important for the
$\gamma$-ray lightcurves if the time scale is sufficiently fast that
the release is significant before the outflow reaches the external
shock. It should not be too fast, however. If the release takes place
before the photospheric radius, i.e. in the optically thick part of
the outflow, the internal energy generated is not radiated away.
Instead, it is converted, through the radial expansion of the shell,
into kinetic energy.

Assume that the outflow is driven by the rotating magnetic field of
the central, compact object, i.e. that the field strength estimates
(\ref{eq:beqr}) or (\ref{eq:bjet}) apply.  Depending on the field
configuration at the source, the field in the flow can be
nonaxisymmetric to a greater or lesser degree, and depending on the
nature of the acceleration process it can be quasi-spherical or more
collimated along the rotation axis. Consider again the two cases
discussed in Sect.~\ref{sec:geo}.

In the perpendicular rotator case, the field consists of `stripes' of
alternating polarity, in which the field energy can be released by
reconnection. In the axisymmetric jet case, the field is unstable to a
kinking process. The magnetic field in both cases is far from a
minimum energy configuration (a potential field). The free energy it
contains is available if it can be released on a sufficiently short
time scale.
  
The initial perturbations from which the MHD instabilities grow are
likely to be present at significant amplitude, from the start of the
outflow, except in the unlikely event that the field configuration of
the source is highly symmetric.  Thus we may assume that the ordered
configurations of Figs.~\ref{fig2} and \ref{fig3} significantly change
to more disordered ones within an MHD instability time scale. These
more disordered configurations then are subject to fast reconnection
processes.
  
Reconnection takes place on time scales governed by the Alfv\'en
speed. It depends on plasma resistivity as well, but in practical
reconnection configurations (as opposed to highly symmetric textbook
examples like tearing mode instability), the resistivity enters only
weakly.  In the Sweet-Parker model for 2-D reconnection, for example
(e.g. Biskamp 1993\nocite{biskamp:93}), it enters through the
logarithm of the magnetic Reynolds number. In more realistic 3-D modes
of reconnection, the basic geometry of reconnection (a `chain link'
kind of configuration) differs from the 2-D geometries. Also, the
reconnection tends to be distributed over many current sheets instead
of a few reconnection points (see Galsgaard \& Nordlund
\cite*{galsgaard:96,galsgaard:97b} for recent numerical results).  The
reconnection rate is still weakly dependent on the resistivity in
these 3-D configurations.

\subsection{Perpendicular rotator}

In the perpendicular rotator, the field in the outflow changes (in the
lab frame) on a length scale $L=\pi c/\Omega$. The time scale
$\tau_\mathrm{r}$ of magnetic energy release scales with the Alfv\'en
crossing time over this length scale. In a comoving frame
\begin{equation}
  1/\tau'_\mathrm{r} \approx \epsilon v'_\mathrm{A}/L',  
\end{equation}
where $\epsilon<1$ is a numerical factor of order unity measuring the
reconnection speed. In the lab frame, the energy release time scale is
$\Gamma$ times longer, hence
\begin{equation}
  \tau_\mathrm{r}=\left\{
    \begin{array}{ll}
      \frac{2\pi\Gamma^2c}{\Omega\epsilon v_\mathrm{A}'}
      &\mbox{(perpendicular rotator)},\\
      \frac{\theta r\Gamma}{\epsilon v_\mathrm{A}'}
      &\mbox{(axisymmetric jet)}.        
    \end{array}\right.
\end{equation}
As the bulk Lorentz factor $\Gamma$ tends to infinity, the energy
release becomes arbitrarily slow. This is also understood by noting
that for $\Gamma\rightarrow\infty$, the electromagnetic field of the
outflow in the observer frame becomes indistinguishable from a pure EM
wave in vacuum, whose energy content is conserved.

The relativistic Alfv\'en speed is (e.g. Jackson
1999\nocite{jackson:99})
\begin{equation}
  v_\mathrm{A}=\frac{c v_B}{(c^2+v_B^2)^{1/2}}, 
\end{equation}
where $v_B=B/(4\pi\rho)^{1/2}$ is the nonrelativistic Alfv\'en speed.

Let the ratio of magnetic energy flux (Poynting flux) to kinetic
energy flux be $\xi$:
\begin{equation}
  \xi = \frac{B^2}{4\pi\Gamma\rho c^2}
  = \frac{B^{\prime 2}}{4\pi\rho' c^2}.
\end{equation}
Hence
\begin{equation}
  v'_\mathrm{A}/c=(1+1/\xi)^{-1/2}. 
\end{equation}
For the models discussed above, $\xi\sim O(1)$.  As long as the
magnetic energy density has not decayed by internal MHD processes, the
comoving Alfv\'en speed is thus of the order of the speed of light.

The magnetic dissipation radius, the distance $r_\mathrm{r}$ from the
source where magnetic energy release becomes important, is thus of the
order
\begin{eqnarray}
  \label{va}
  r_\mathrm{r}&=&c\tau_\mathrm{r}
  =\frac{\pi c}{\epsilon\Omega}\Gamma^2 (1+1/\xi)^{1/2}
  \quad\mbox{(perp. rotator)}\\
  &=&2\cdot10^{12}\,\mathrm{cm}\cdot\Omega_4^{-1}
  \left(\frac{\Gamma}{300}\right)^2
  \frac{1}{\epsilon_{-1}}
  (1+1/\xi)^{1/2}.
\end{eqnarray}

The length scale in the comoving frame, $L'$, is equal to the
wavelength $\lambda' = 2\pi c\Gamma/\Omega = 2\pi r_\mathrm{L}$ in the
perpendicular rotator case) or the lateral scale, $L'=\theta r$ (in a
collimated outflow).

\subsection{Axisymmetric jet}

For the case of an axisymmetric outflow along the rotation axis, the
situation is a bit different. The length scale $L$ of the magnetic
field is now the jet radius, $L\approx\theta r$, where $\theta$ is the
opening angle of the jet. Since this is measured perpendicular to the
flow, it is the same in the lab and the comoving frame. In the
comoving frame, the time for an Alfv\'en wave to communicate over this
distance is $\tau'_\mathrm{A} = r\theta/v'_\mathrm{A}$, in the lab
frame $\tau_\mathrm{A} = \Gamma r\theta/v_\mathrm{A}$. This is less
than the time for the flow to reach a distance $r$, $t = r/\beta
c\approx r/c$ if
\begin{equation}
  \theta < \frac{v'_\mathrm{A}}{\Gamma c}
  = \frac{1}{\Gamma(1+1/\xi)^{1/2}}.  
\end{equation}
If the jet is wider than this, parts of the flow moving at angles
separated by more than $\vartheta =(1+1/\xi)^{-1/2}/\Gamma<1/\Gamma$ have no
time to communicate by an Alfv\'en wave in the time elapsed since the
start of the flow, and behave as if they are causally disconnected
from each other. Kink instability will thus be limited to an inner
core of opening angle $\vartheta<1/\Gamma$.  In directions outside
this angle, the instability is suppressed since it has to communicate
from one side to then other across the axis of the jet.  This reduces
the fraction of the available magnetic energy that can be tapped by
MHD processes.

On the other hand, magnetic energy release in the inner core so
defined already starts very close to the source (where the length
scale $\vartheta r$ is small). Thus only a fraction of the dissipation
will happen in the optically thin, observable regions.

\subsection{General nonaxisymmetric outflows}

The conclusion from the above is that in the case of a perpendicular
rotator, there is a well defined `magnetic dissipation radius'
$r_\mathrm{r} \sim 10^{12}$\,cm where most of the magnetic energy is
dissipated. For a purely axisymmetric outflow along the rotation axis,
on the other hand, only a fraction of the magnetic energy can be
released, unless the opening angle is less than $\sim 1/\Gamma$. This
fraction probably dissipates close to the source, and not necessarily
in the optically thin region where it could contribute to the observed
emission. In intermediate cases, where both an axisymmetric and a
nonaxisymmetric component are present, the magnetic field in the
outflow changes direction on the length scale $L = \pi c/\Omega$,
without completely changing direction. In such cases, the amount of
the magnetic energy that can be released in directions outside the
central core $\vartheta \sim 1/\Gamma$ is of the order
$B_\mathrm{n}^2/8\pi$, where $B_\mathrm{n}$ is the nonaxisymmetric
part of the field. This is a siginificant fraction of the total
magnetic energy unless the field is nearly axisymetric.  The
axixymmetric jet case, though attractive as a computable model, is
thus rather singular with respect to the question of magnetic
dissipation which we address in this paper.


\section{Photospheric radius}

The reconnection radii derived above need to be compared with the
radius $r_\mathrm{p}$ of photosphere in the outflow. If $r_\mathrm{r}$
is larger than $r_\mathrm{p}$, the dissipation of magnetic energy
takes place mainly in the optically thin regime, and the dissipated
energy is radiated away locally. On the other hand if $r_\mathrm{r} <
r_\mathrm{p}$, the energy released from the initially ordered field
configuration increases the internal energy of the plasma. The radial
expansion of the flow converts this energy into kinetic energy of
outflow. Though this may be useful in obtaining large Lorentz factors,
it also implies that the magnetic energy left in the flow by the time
it passes through the photosphere is small if $r_\mathrm{r} \ll
r_\mathrm{p}$. The net output of a magnetically driven fireball with
magnetic dissipation taking place mostly in the optically thick regime
would be essentially the same as a standard non-magnetic fireball,
with the attendant problem of a low efficiency of the production of
radiation by internal shocks.

The photospheric radius in a steady relativistic outflow has been
derived by Abramowicz et~al. \cite*{abramowicz:91}. The optical depth
of a moving shell of constant (in time) density is a Lorentz
invariant, hence the same for photons moving along the direction of
the flow and those moving in the opposite direction (assuming also
that the opacity is energy-independent). In a radial outflow, however,
a photon sees a different history of mass density depending on its
direction.

In addition to this effect, we take into account the finite duration
$\tau$ of the outflow. Assume a total amount of mass $M$ is ejected
with a constant Lorentz factor $\Gamma$, at a rate $\dot M$ which is
is constant in the time interval $0<t<\tau$ (and zero outside this
interval).  If at some time after the end of the outflow the inner
edge of the outward moving shell of mass is at a radius
$r_\mathrm{i}$, the optical depth $\tau_\mathrm{i}$ of the shell as
seen by a photon propagating outward from this radius is
\cite{abramowicz:91}:
\begin{equation}
  \label{eq:ph}
  \tau_\mathrm{i} = 
  \int_{r_\mathrm{i}}^{r_\mathrm{i}+c\tau/(1-\beta)} 
  \rho' \kappa \Gamma(1-\beta)\,\mathrm{d}r,
\end{equation}
where $\rho'$ is the rest mass density in the comoving frame. To
lowest order in $\Gamma^{-1}$, $\rho'=M/(4\pi r^2 c\tau\Gamma)$, and
we find
\begin{equation}
  \tau_\mathrm{i}= \frac{M\kappa}
  {4\pi r_\mathrm{i} \left[r_\mathrm{i}+c \tau/(1-\beta)\right]}.
\end{equation}
For very small duration $\tau$, this reduces to the optical depth
$M\kappa/(4\pi r_\mathrm{i}^2)$ for a shell of mass $M$ and fixed
radius $r_\mathrm{i}$.

The optical depth is a decreasing function of time as the shell moves
outward and $r_\mathrm{i}$ increases. Define now the photospheric
radius $r_\mathrm{p}$ as the value of $r_\mathrm{i}$ for which
$\tau_\mathrm{i}=1$.
With $1/(1-\beta)\approx 2\Gamma^2$ and solving
for 
\begin{equation}
  r_\mathrm{i} = c\tau\Gamma
  \left[
    \left(1+\frac{M\kappa}{4\pi (c\tau\Gamma^2)^2}\right)^{1/2}-1
  \right],
\end{equation}
we encounter the parameter 
\begin{equation}
  \frac{M\kappa}{4\pi (c\tau\Gamma^2)^2} = 
  1.82\cdot 10^{-5}\cdot
  \frac{E_{52}}{\xi+1}
  \parth{\tau}{3}{s}^{-2}
  \left(\frac{\Gamma}{300}\right)^{-5}.
\end{equation}
Thus, for relevant GRB parameters, we find the photospheric radius to
be
\begin{equation}
  \label{eq:rp}
  r_\mathrm{p}=\frac{M\kappa}{8\pi c\tau\Gamma^2}  
  =\frac{E\kappa}{8\pi c^3\tau\Gamma^3 (\xi+1)}.
\end{equation}
Numerically,
\begin{equation}
  r_\mathrm{p}=7.2\cdot10^{10}\,\mathrm{cm}\cdot
  (\xi+1)^{-1}
  E_{52}
  \parth{\tau}{3}{s}^{-1}
  \left(\frac{\Gamma}{300}\right)^{-2}. 
\end{equation}
As soon as the inner edge of the shell has expanded to this radius,
radiation can escape from the entire shell. Due to the relativistic
effect noted by Abramowicz et~al. \cite*{abramowicz:91}, the
photospheric radius (\ref{eq:rp}) is much smaller than if computed for
a shell of fixed radius.

\subsection{Pair opacity}

In addition to the baryons, pairs could also contribute to the
opacity.  The photospheric radius (\ref{eq:rp}) then does not apply,
because it includes only the constant opacity of scattering on the
electrons associated with the baryonic mass.

%
Assume a steady wind in which pairs dominate the kinetic energy flux
and the opacity, and in which the kinetic energy of the pairs is a
fraction $\xi_\pm$ of the total luminosity $E/\tau$.
Let $T\approx 2\cdot10^8$\,K be the temperature at the pair
photosphere (due to the steep dependence of the pair density on
temperature, this value does not depend much on conditions in the
outflow).  Ignoring the opacity due to baryonic matter, the radius of
the pair photosphere $r_\pm$ is \cite{usov:94}:
\begin{eqnarray}
  \label{eq:rpairs}
  r_{\pm} &=&
  \left(\frac{\xi_\pm E/\tau}{4\pi\sigma_{\mathrm SB} T^4\Gamma^2}
  \right)^{1/2}\\
  &=&
  2\cdot10^8\,\mathrm{cm}\cdot
  (\xi_\pm E_{52})^{1/2}\nonumber\\
  &&
  \cdot
  \parth{T}{2\cdot10^8}{K}^{-2}
  \left(\frac{\Gamma}{300}\right)^{-1}
  \parth{\tau}{3}{s}^{-1/2}.
\end{eqnarray}
where $\sigma_\mathrm{SB}$ is the Stefan-Boltzmann constant and
$\xi_\pm$ is the energy fraction in the pairs.

This value is much smaller than the photospheric radius (\ref{eq:rp}).
Thus, for typical baryon-loaded GRB parameters pairs annihilate before
they reach the optically thin domain.


\section{Discussion}
\label{sec:disc}

Magnetic fields may well be the main agent tapping the
rotational/gravitational energy in the central engines of GRB (for
reviews see M\'esz\'aros \cite*{meszaros:99b}).
Alternatives like the production of pair plasma fireballs by neutrino
annihilation have turned out to have a rather small efficiency of
conversion of gravitational energy \cite{ruffert:97}.  A fireball
powered by magnetic fields (`magic hydrodynamics') can in principle
produce $\gamma$-rays in the same way as in field-free mechanisms,
namely by internal shocks in the optically thin part of the flow. A
well-known problem with the internal shock mechanism, however, is the
low efficiency of conversion to gamma-rays \cite{daigne:98}), of the
order of a few percent (see however Kumar \& Piran
2000\nocite{kumar:00}).  A magnetically powered outflow naturally
carries a magnetic field with it (`Poynting flux'). This raises the
question whether dissipation of this internal magnetic energy in the
outflow can perhaps produce the observed radiation with a better
efficiency.

We have addressed this question by considering a few possible
scenarios for magnetic fields in GRB outflows. In the first, the
`passive scenario', the magnetic field is assumed not to be
responsible at all for powering the outflow, but only advected
passively by an outflow produced by something else. We find that the
maximum field strengths possible in this case are small compared with
equipartition with the kinetic energy of the outflow, but still
potentially significant for the effective production of synchrotron
and/or inverse Compton emission in the internal shock model.

As magnetically-driven models we consider the case of a quasispherical
outflow produced by a rotating non-axisymmetric magnetic field, and
the case of a jet-like outflow along the axis of a rotating
axisymmetric field. The quasispherical case is like the models
produced for the Crab pulsar \cite{coroniti:90,gallant:94}.  It has
been developed for a completely baryon-free, pure pair-plasma outflow
by Usov \cite*{usov:94}. The jet case is similar to
magnetohydrodynamically driven wind models
\cite{blandford:82,sakurai:85}.

The magnetic field in all these cases is confined in an outward moving
shell of width $c\tau$, where $\tau$ is the duration of the flow. If
internal dissipation is ignored in the outflow, the total magnetic
energy in the shell is constant, and the field strength $B$ varies
with distance as $r^{-1}$ ($B$ and $\tau$ measured in the observer
frame). The configuration of the magnetic field in the shell is
different in each case (see figures \ref{fig1}--\ref{fig3}). If the
central engine is time-dependent, for example in the form of a series
of sub-bursts, magnetic shells like this are produced by each
sub-burst, and the magnetic flux and energy in the shells is renewed
with each burst (i.e. not limited by the magnetic flux of the central
object).

We find that the MHD approximation (in the sense that the electric
field in the fluid frame is small compared with the magnetic field) is
safe out to large distances from the source, of the order $10{19}$\,cm
or more. This is due to the relatively large amount of baryons in the
outflow compared with, say, a pulsar wind situation. This simplifies
the discussion of magnetic energy dissipation at least conceptually,
since the energy release can be studied without detailed discussions
of plasma processes (though they may enter again in the discussion of
what produces the radiation).

The field strength in a magnetically driven outflow depends on the
extent to which internal MHD processes have been able to dissipate
magnetic free energy in the flow. In the absence of such processes,
the magnetic energy density is typically of the same order as the
kinetic energy density in magnetically driven flows, and the field is
well-ordered (large scale). In the internal shock model, such fields
are sufficient to produce synchrotron and/or inverse Compton emission
in the gamma-ray range. The synchotron case is favoured by a higher
efficiency during the whole course of the burst.  There are now some
observational evidences that the magnetic field required for the
afterglow emission represents a small fraction of the equipartition
value whereas the prompt gamma-ray emission via internal shocks is
possible only with more intense fields close to the equipartition
value \cite{galama:99}. In the standard picture where the afterglow is
due to the forward shock propagating in the external medium where no
large scale field is present, it could mean that the generation of a
local magnetic field behind the shock wave is inefficient. It is then
possible that this locally generated field is also very small behind
internal shocks within the shell but we show here that the large-scale
field present in a magnetically driven outflow has no difficulty in
most cases to supply the strength necessary for synchrotron and/or
inverse Compton emission.

The rate of energy release through MHD processes like instabilities
and fast reconnection is governed by the Alfv\'en speed. In the rest
frame of the outflow, the Alfv\'en speed is of the order of the speed
of light if the kinetic and magnetic energy fluxes are similar.  In
outflows produced by a perpendicular rotator, the magnetic field
changes sign on the small length scale $c/\Omega$, and most of the
magnetic energy can be dissipated by reconnection.  The typical radius
at which the energy release takes place in this case is $\sim
10^{12}$\,cm for standard GRB parameters ($10^{52}$\,erg, $3$\,s,
$\Gamma=300$). The photospheric radius, on the other hand, is small
(mostly because of the relativistic effect discussed by Abramowicz
et~al. \cite*{abramowicz:91}), $\sim 10^{10}$\,cm.  We can thus be
confident that dissipation of energy stored in the magnetic field of
the outflow occurs in the optically thin regime.

In the case of an outflow produced by a purely axisymmetric rotating
magnetic object such changes of sign do not occur. Energy can still be
released by kink instabilities, but for causality reasons these are
not effective outside a narrow cone of angle $1/\Gamma$ near the
rotation axis (section \ref{sec:decay}).  The expected amount of
magnetic dissipation in the optically thin regime is quite limited for
such an axisymmetric field.  A purely axisymmetric configuration,
however, is a singular case, a priori unlikely for magnetic fields
produced in a transient object like the central engine of a GRB. In
the more likely case that a nonaxisymmetric field component is present
as well, the energy in this component can be released by reconnection
in nearly the same way as in the case of a perpendicular rotator.

Magnetic energy dissipation in the optically thin regime is probably
not as simple as the shock dissipation in the internal shock model. In
particular, the way in which nonthermal electron distributions are
produced still needs to be investigated. It is likely that nonthermal
radiation is again produced, as in the internal shock model, but the
shape of the radiation spectrum may be more difficult to compute, as
well as the typical time scale of the radiative process, which should
be short compared to the expansion time scale (see (\ref{eq:tex}):
$t'_\mathrm{ex} \sim 1\,\mathrm{s}\cdot ({\Gamma}/{300})^{-1}$ at
$10^{13}$\,cm) to have a high radiative efficiency.

The main attraction of of GRB radiation produced by magnetic dissipation 
in a magnetically driven
outflow is the efficiency with which the energy flux from the central
engine can be converted into observable radiation. This efficiency is
limited only by the ratio of Poynting flux to total energy flux in the
flow. In the magnetically driven cases considered, this ratio can be
close to unity. In the internal shock scenario, on the other hand,
only a fraction $\la 0.1$ of the energy is dissipated in the optically
thin region. Since the bulk kinetic energy of the burst is dissipated
in the afterglow, the internal shock model predicts the afterglows to
dominate the energy output, which is probably inconsistent with
current observations. In magnetic dissipation models such as those
discussed here, the energy
emitted in the afterglow can in principle be arbitarily small compared
with the prompt emission. It is not yet clear, however, how much of
the magnetically dissipated energy can be in the form of
$\gamma$-rays.


\bibliography{aamnem99,mhdgrb}
\bibliographystyle{aabib99}

\end{document}